\documentclass[aps,prl,reprint,twocolumn,superscriptaddress,raggedbottom,floatfix,nofootinbib]{revtex4-2}
\usepackage[normalem]{ulem}
\usepackage{amsmath,amssymb,amsthm}
\usepackage{booktabs,xcolor}
\usepackage{bm}
\usepackage{mathtools}
\usepackage{graphicx}
\usepackage[hidelinks]{hyperref}
\hypersetup{pdftitle={A Bound on Wormhole Probing Time from Holographic Multi-Entropy},
 pdfauthor={Takanori Anegawa and Norihiro Iizuka}}

\NewDocumentCommand{\GM}{o}{%
  \mathrm{GM}\IfValueT{#1}{^{(#1)}}%
}

\newcommand{\rp}{r_+}
\newcommand{\Lc}{\mathcal L}
\newcommand{\Lhat}{\widehat{\mathcal L}}

\begin{document}
\setcounter{secnumdepth}{2}

\title{A Bound on Wormhole Probing Time from Holographic Multi-Entropy}

\author{Takanori Anegawa}
\affiliation{Yonago College, National Institute of Technology,
Yonago, Tottori 683-8502, Japan}

\author{Norihiro Iizuka}
\affiliation{Department of Physics, National Tsing Hua University, Hsinchu 300044, Taiwan}
\affiliation{Yukawa Institute for Theoretical Physics, Kyoto University, Kyoto 606-8502, Japan}

\date{\today}

\begin{abstract}
Bipartite entanglement entropy eventually saturates and therefore cannot track the continued growth of an eternal wormhole. Recently, it was shown that genuine tripartite multi-entropy can probe wormhole growth even after bipartite entanglement entropies have saturated. This naturally raises the question: can increasing the number of parties extend the probing time indefinitely? Adopting the multiway maximin prescription and assuming that its selected representatives are regular spacetime-stationary networks, we prove that the answer is no for nonrotating BTZ black holes. We establish a common finite upper bound on the probing time for every fixed finite boundary partition, independent of the number, sizes, and arrangement of its parties. For any network on any Cauchy slice, there exists a comparison network that assigns the entire central region containing the wormhole interior to a single party. A network longer than this comparison network cannot minimize length on that slice. The central length required by a stationary network connecting the two boundaries grows with time, whereas the central length of the comparison network remains uniformly bounded. At sufficiently late times, such a network therefore cannot minimize length on any Cauchy slice containing it and is excluded from the maximin prescription. Consequently, beyond this bound, no network selected by maximin enters the black-hole interior.
\end{abstract}

\maketitle

\section{Introduction}
\label{sec:introduction}
Holographic entanglement entropy relates boundary quantum information to bulk geometry through the Ryu--Takayanagi (RT) prescription~\cite{Ryu:2006ef}. 
The $\mathtt q$-party multi-entropy $S^{(\mathtt q)}$ is the $n\to1$ limit of the corresponding multipartite Rényi construction and reduces to ordinary entanglement entropy for $\mathtt q=2$~\cite{Gadde:2022cqi}. Its proposed classical holographic dual is the area of a minimal multiway cut: a network of interfaces partitioning the bulk into regions homologous to the prescribed boundary parties. Such networks can probe geometric information different from bipartite RT surface data. Several studies suggest that the deep infrared (IR) region \cite{Iizuka:2025ioc,Iizuka:2025bcc,Iizuka:2025caq,Ju:2025eyn,Iizuka:2026qqg,Iizuka:2026ahd,Iizuka:2026kbo,Anegawa:2026npk,Fujiki:2026ucr} is associated with multipartite entanglement. Related questions have also been explored in the context of the Markov gap \cite{Akers:2019gcv,Hayden:2021gno}.

The growing Einstein–Rosen (ER) bridge of an eternal black hole provides a dynamical setting for testing how deep such networks can probe. Under equal future-directed evolution of the two boundaries, the bridge continues to grow, whereas bipartite entanglement entropies for finite boundary regions saturate on timescales set by their sizes~\cite{Hartman:2013qma}. This limitation motivated the complexity–volume proposal~\cite{Susskind:2014rva, Stanford:2014jda}.

%\begin{figure*}[t!]
%\centering
%\includegraphics[width=\textwidth]{figures/wormhole_partitions.pdf}
%\caption{The question for arbitrary finite partitions. The same fixed boundary labels are shown at two times: the left boundary has two unequal intervals and the right has three, giving an odd party number $\mathtt q=5$. 
%The two horizon sections enclose the growing interior. The theorem allows any finite number and arrangement of intervals at $t_L=t_R=t$, under the stated assumptions. 
%This is a schematic spatial bridge, not an isometric embedding or a plot of the interior length.}
%\label{fig:wormhole-partitions}
%\end{figure*}

Recently, Ref.~\cite{Anegawa:2026yee} showed that genuine tripartite multi-entropy can detect continued wormhole growth after all bipartite entropies of the specified parties have saturated, although it eventually saturates as well. Here we ask whether increasing the number of parties or choosing their sizes and arrangement appropriately can extend the sensitivity to dynamical wormhole growth to arbitrarily late times.\footnote{ Closely related works~\cite{Iizuka:2026kbo,Anegawa:2026npk} demonstrate that genuine quadripartite multi-entropy can detect continuous moduli of static wormholes that are invisible to all bipartite and tripartite multi-entropies.}

Under the variational assumptions specified below, we show that the answer is no within the classical nonrotating BTZ setting. 
More precisely, for any fixed finite $\mathtt q$-partition $\mathcal P$ of the boundary, the corresponding leading classical holographic multi-entropy $S^{(\mathtt q)}$ cannot probe the black-hole interior at times $t>t_{\rm U}(r_+)$, where $t_{\rm U}(r_+)$ is a common finite upper bound independent of the number $\mathtt q$, sizes, and arrangement of the parties.

The proof does not require classifying all network configurations.
Any network on any Cauchy slice admits a comparison network that assigns the central region containing the wormhole interior to a single party, leaving the partition outside unchanged.
This removes all central interfaces and adds only a bounded interface length along two exterior cut circles, independently of time and party number.
Spacetime stationarity and causality imply that networks connecting the two boundaries require a central interface length whose lower bound grows with time.
Eventually, the comparison network is shorter on every Cauchy slice containing such a network, excluding such stationary networks from the maximin prescription.

A holographic prescription for multi-entropy in time-dependent geometries has not yet been established. We therefore adopt the multiway maximin prescription as a working assumption.
In this prescription, one first minimizes the total interface length over all admissible bulk partitions on each Cauchy slice, and then maximizes this minimum over the slices. The first minimization does not require geodesic edges or balanced junctions. 

In addition, we assume that the network selected by multiway maximin is a regular spacetime-stationary network, {\it i.e.}, it consists of finitely many spacelike geodesic edges satisfying the junction balance condition. 
This is motivated by ordinary holographic entanglement entropy, for which stable maximin surfaces are known to be HRT extremal surfaces~\cite{Wall:2012uf}.
The analogous property for multiway networks is assumed here.

\section{Setup and the recoloring bound}
\label{sec:setup}
\subsection{Geometry, parties, and the observable}
We use eternal nonrotating BTZ with AdS radius one:
\begin{equation}
 \begin{gathered}
 ds^2=-(r^2-\rp^2)dt_\sigma^2+
 \frac{dr^2}{r^2-\rp^2}+r^2d\phi^2,\\
 \phi\sim\phi+2\pi,\qquad \rp>0.
 \end{gathered}
 \label{eq:metric}
\end{equation}
The sign $\sigma=-1$ labels the left exterior and $\sigma=+1$ the
right exterior. Each $t_\sigma$ increases toward the future in its
own exterior. We set
\begin{equation}
 t_L=t_R=t\ge0.
\end{equation}
These equal future-directed translations are not one global Killing
translation, whose future orientations are opposite at the two ends. 
We consider a fixed partition $\mathcal P$ of the two boundary circles into $\mathtt q\ge2$ nonempty parties.
Each party may be disconnected and may occupy portions of either or both boundaries.
We allow finite unions of intervals as well as whole boundary circles, with no restrictions on their sizes or arrangement; in particular, no left-right symmetry is required.
The angular endpoints and party labels are held fixed as $t$ varies.

Let $\Lc(N)$ denote the total length of a network $N$, counting each interface once.
For each Cauchy slice $\Sigma$ anchored at $t_L=t_R=t$, we consider all partitions into labeled regions that agree with $\mathcal P$ at the boundary. Each labeled region may be disconnected. The interfaces may meet at junctions in the bulk.
We minimize the total interface length over the resulting
networks $N$, and then maximize this minimum over $\Sigma$:
\begin{equation}
 S^{(\mathtt q)}(t)
 =\frac{1}{4G_N}\max_{\Sigma}\min_N\Lc(N).
\end{equation}
We suppress the dependence on the fixed boundary partition $\mathcal P$. 
The inner minimization includes non-stationary networks; geodesic edges and junction balance are assumed only for the selected maximin representative.

We call an edge extending to the asymptotic boundary a boundary leg.
The comparisons below leave the asymptotic portions of all boundary legs unchanged, so their UV divergences cancel.
The subtraction convention is given in Appendix~\ref{app:embedding}.

\subsection{Recoloring the central region}
\label{sec:recoloring}
Let $\Sigma_*$ be a Cauchy slice of BTZ annulus topology anchored at $t_L=t_R=t$, and let $N_*$ globally minimize interface length on $\Sigma_*$.
Choose a dimensionless constant $R>1$, held fixed as $t$ varies, and define two cut circles $C_L,C_R$ at $r=R\rp$,
one in each exterior. We call the region between them the \emph{central region} $B$.
It contains the wormhole interior, $0<r<\rp$, on this slice, together with adjacent portions of both exteriors.
The total interface length of $N_*$ within $B$ is
\begin{equation}
 D_B=\Lc(N_*\cap B).
 \label{eq:central-definitions}
\end{equation}

\begin{figure*}[t!]
\centering
\includegraphics[width=\textwidth]{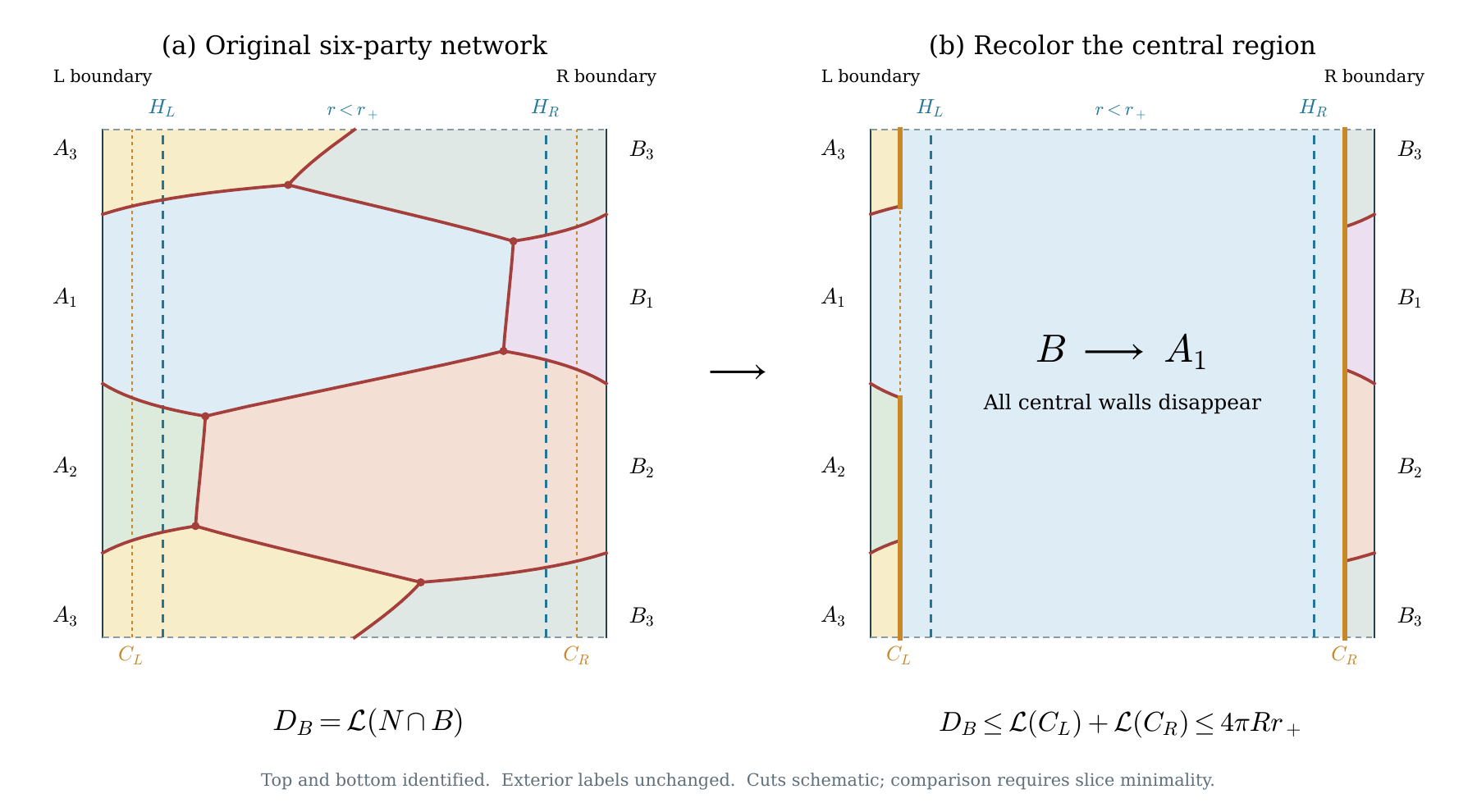}
\caption{Recoloring the central region. Top and bottom are identified. 
(a) An illustrative six-party network.
Blue dashed lines mark the future horizons, and the exterior
cuts $C_L,C_R$ are shown schematically.
(b) Assigning label $A_1$ throughout $B$ removes all central
interfaces while preserving the partition outside $B$.
Orange segments show the new interfaces along the cuts.
For a slice-minimizing network, this comparison gives
$D_B\le\mathcal L(C_L)+\mathcal L(C_R)\le4\pi Rr_+$. The recolored comparison network need not be stationary.
}
\label{fig:recoloring-bound}
\end{figure*}

We now assign a single existing party label throughout $B$, leaving the partition outside $B$ unchanged, as illustrated in Fig.~\ref{fig:recoloring-bound}.
Denote the resulting network by $N_{\rm rec}$.
The recolored network corresponds to an allowed bulk partition with the same boundary partition $\mathcal P$ and is therefore a valid comparison network in the maximin prescription.
All interfaces within $B$ disappear, and new interfaces can arise only along $C_L$ and $C_R$.
The total added length is therefore at most $\Lc(C_L)+\Lc(C_R)$.

The Cauchy property implies that each timelike line at fixed
$(r,\phi)$ in an exterior meets $\Sigma_*$ once.
Thus each cut circle $C_\sigma$ can be parametrized as
\begin{equation}
(r,t_\sigma,\phi)=(R\rp,t_\sigma(\phi),\phi),
\end{equation}
where $\sigma=-1$ labels the left exterior and
$\sigma=+1$ the right exterior. The exterior metric gives
\begin{equation}
\begin{aligned}
 \Lc(C_\sigma)
 &=\rp\int_0^{2\pi}
 \sqrt{R^2-(R^2-1)(\partial_\phi t_\sigma)^2}\,d\phi\\
 &\le2\pi R\rp.
\end{aligned}
 \label{eq:cut-bound}
\end{equation}
Achronality makes the integrand real, and it is then bounded by $R$.
Since $N_*$ minimizes length on $\Sigma_*$, this replacement
cannot shorten it. Therefore, $N_*$ must satisfy
\begin{equation}
 D_B\le\Lc(C_L)+\Lc(C_R)\le4\pi R\rp.
 \label{eq:central-upper}
\end{equation}

\section{A uniform bound on interior connections}
\label{sec:proof}
We now consider stationary networks with a connected component joining the two asymptotic boundaries.
We derive a lower bound on their interface length within $B$ that grows with boundary time.
Comparing this with the upper bound~\eqref{eq:central-upper} shows that such networks cannot be selected by maximin at sufficiently late times.

\subsection{Causality and the length of a surviving connection}

Consider the Cauchy slice $\Sigma_*$ anchored on both boundary circles at $t_L=t_R=t$. Since the cut circles $C_L,\ C_R$ lie on this same achronal slice, the time difference between a cut point and the boundary point at the same angle cannot exceed the radial light-travel time between them. A radial null ray at fixed angle takes exterior time
\begin{equation}
 \int_{R\rp}^{\infty}\frac{dr}{r^2-\rp^2}
 =\frac{a_R}{\rp}, \qquad a_R=\operatorname{arctanh}\frac1R
\label{eq:aR}
\end{equation}
to reach the boundary. 
Since both points lie on the same achronal slice, they must be spacelike or null separated.
Achronality therefore gives, at every point of either cut,
\begin{equation}
 |\rp t_\sigma(\phi)-T|\le a_R,\quad T=\rp t.
 \label{eq:causal-bound}
\end{equation}

We describe BTZ using the AdS$_3$ embedding given in Appendix~\ref{app:embedding}:
\begin{equation}
\begin{aligned}
 X^0&=\rho\cosh p,
 &X^1&=\sqrt{\rho^2-1}\sinh(\rp t_\sigma),\\
 X^2&=\rho\sinh p,
% X^1&=\sqrt{\rho^2-1}\sinh(\rp t_\sigma),
 &X^3&=\sigma\sqrt{\rho^2-1}\cosh(\rp t_\sigma),
\end{aligned}
\end{equation}
where $\rho=r/\rp$, $p=\rp\phi$, and $\sigma=-1$ and $+1$ label the left and right exteriors, respectively.

The angular identification $\phi\sim\phi+2\pi$ changes $X^0$ and $X^2$, but leaves $X^1$ and $X^3$ unchanged. Thus $X^1$ and $X^3$ are single-valued functions on BTZ, even along paths that wind around the angular direction. They also capture the two features needed here: $X^1$ increases with exterior time at fixed radius, while $X^3$ has opposite signs in the two exteriors. We therefore focus on
\begin{equation}
 Y=X^1,\qquad W=X^3.
\end{equation}

At every point on $C_L$ and $C_R$, we have $\rho=R$,
and the causal bound~\eqref{eq:causal-bound} gives
$\rp t_\sigma(\phi)\ge T-a_R$.
Since $\sinh$ is increasing, on $C_L \cup C_R$, we obtain
\begin{align}
 Y&\ge\sqrt{R^2-1}\sinh(T-a_R)\nonumber\\
  &=R\sinh T-\cosh T\nonumber\\
  & \equiv Y_{\mathrm{bound}}(R,T).\qquad 
(\text{on } C_L\cup C_R)
 \label{eq:boundary-H}
\end{align}
Here we used $\cosh a_R=R/\sqrt{R^2-1}$ and $\sinh a_R=1/\sqrt{R^2-1}$, which follow from Eq.~\eqref{eq:aR}.

The quantity $Y_{\mathrm{bound}}(R,T)$ is a guaranteed
lower bound on $Y=X^1$ at every point of $C_L\cup C_R$,
not necessarily the actual minimum of $Y$ on these circles.
This lower bound on $Y$ at both cut circles grows with the boundary time. Meanwhile, $\cosh(\rp t_\sigma)>0$ implies $W<0$ everywhere on $C_L$ and $W>0$ everywhere on $C_R$. Any continuous path connecting the two cut circles must therefore pass through a point where $W=0$.

For a spacelike geodesic $X(s)$ on the AdS hyperboloid, parametrized by its proper length $s$, we have
$X\cdot X=-1$, $X'\cdot X'=1$, 
where primes denote derivatives with respect to $s$. Differentiating the first constraint twice gives
\begin{equation}
 X\cdot X'=0,\qquad
 X\cdot X''=-X'\cdot X'=-1.
\end{equation}

Since the curve is a geodesic, its acceleration $X''$ in the embedding space is normal to the AdS hyperboloid. The normal direction is given by $X$ itself, so $X''=\lambda X$ (see Appendix~\ref{app:embedding}). Taking the inner product with $X$ gives $-1=-\lambda$, hence
\begin{equation}
 X''=X.
\end{equation}
Taking the $X^1$ and $X^3$ components gives $Y''=Y$ and $W''=W$. These equations follow from the geodesic condition and are not additional assumptions about the network.

At each junction of bulk interfaces, stationarity under
variations of its position requires equal-tension force balance:
\begin{equation}
 \sum_{e\ni v}\tau_{e,v}^{\mu}=0,
 \label{eq:balance}
\end{equation}
where $\tau_{e,v}^{\mu}$ is the unit tangent directed from the vertex $v$ into the incident edge $e$. Hence the derivatives of each embedding component, directed into the incident edges, sum to zero. 
Combining the geodesic equations with the junction balance conditions projected onto $Y$ and $W$ gives
\begin{equation}
\begin{gathered}
 Y''=Y,\qquad W''=W,\\
 \sum_{e\ni v}\partial_{n_e}Y=0,\qquad
 \sum_{e\ni v}\partial_{n_e}W=0.
\end{gathered}
 \label{eq:graph-equations}
\end{equation}
Hereafter, we call the conditions in the second line the Kirchhoff conditions (or junction balance conditions) for $Y$ and $W$.

\subsection{Why branches and cycles cannot evade the bound}
\label{sec:branched-length}
If $N_*$ is selected by maximin, it satisfies Eq.~\eqref{eq:graph-equations} by our stationarity assumption. 
Suppose a connected component of $N_*$ attaches to both asymptotic boundaries. Choose a simple path in that component from a left boundary leg to a right boundary leg, with endpoints outside the cuts. Traverse it from left to right and stop at its first intersection with $C_R$.
Take the last intersection with $C_L$ before that point. 
The segment between these intersections defines a simple path $\gamma$ contained in $B$, even if the original path crosses either cut circle repeatedly.

Let $\Gamma$ be the connected component of $N_*\cap B$ containing $\gamma$, including all branches and cycles belonging to that component. Thus,
\begin{equation}
\gamma \subseteq \Gamma \subseteq N_* \cap B \subseteq N_*.
\end{equation}
For the illustrative network in Fig.~\ref{fig:recoloring-bound}(a), the intersection with $B$ is connected, so $\Gamma$ contains all interfaces within $B$.
We derive a lower bound on the total length of $\Gamma$, including all its branches and cycles. 
The Kirchhoff conditions remain valid at its junctions inside $B$, while its boundary vertices on $C_L\cup C_R$ satisfy $Y\ge Y_{\mathrm{bound}}(R,T)$.

We now restrict to $Y_{\mathrm{bound}}(R,T)>1$. The minimum argument in Appendix~\ref{app:graph} gives $Y>0$ throughout $\Gamma$. Recall that $Y\ge Y_{\mathrm{bound}}(R,T)>1$
at both endpoints of the path $\gamma$,
which lie on the cut circles $C_L$ and $C_R$, respectively.
Since $W$ has opposite signs at these endpoints,
continuity implies that there exists a point
$x_0\in\gamma$ where $W(x_0)=0$.
The embedding constraint
\begin{equation}
 Y^2-W^2=1-\rho^2
\end{equation}
then gives
\begin{equation}
 0<Y(x_0)\le1.
\end{equation}
Thus, along $\gamma$, $Y$ must decrease from $Y\ge Y_{\mathrm{bound}}(R,T) >1$ to $Y(x_0)\le 1$ at $x_0$,
and then increase again to $Y\ge Y_{\mathrm{bound}}(R,T)$.

To understand the resulting length bound, first consider the simple case where $\Gamma$ is a single geodesic
without junctions.
Along this geodesic, $Y''=Y$.
If $s=0$ is the point where $Y$ reaches its minimum
$1 \ge m > 0$, the solution is
\begin{equation}
 Y(s)=m\cosh s.
\end{equation}
Therefore, reaching $Y\ge Y_{\mathrm{bound}}(R,T)$ from the minimum requires a proper length of at least $\operatorname{arccosh}(Y_{\mathrm{bound}}(R,T)/m)\ge\log Y_{\mathrm{bound}}(R,T)$ on each side. 
The total length is consequently bounded below by $2\log Y_{\mathrm{bound}}(R,T)$.

For a network with junctions, the derivatives of $Y$ along $\gamma$ need not be continuous at the junctions.
Nevertheless, the Kirchhoff conditions imply a logarithmic lower bound on the total length of the connected component $\Gamma$, including all branches and cycles, as proved in Appendix~\ref{app:graph}:
\begin{equation}
 \Lc(\Gamma)\ge2\log Y_{\mathrm{bound}}(R,T).
 \label{eq:graph-lemma}
\end{equation}
For fixed $R>1$, the lower bound grows as $2\log Y_{\mathrm{bound}}(R,T)=2\rp t+O(1)$, with the same leading rate as a two-sided BTZ geodesic~\cite{Stanford:2014jda}.

\begin{figure}[tbp]
\centering
\includegraphics[width=\columnwidth]{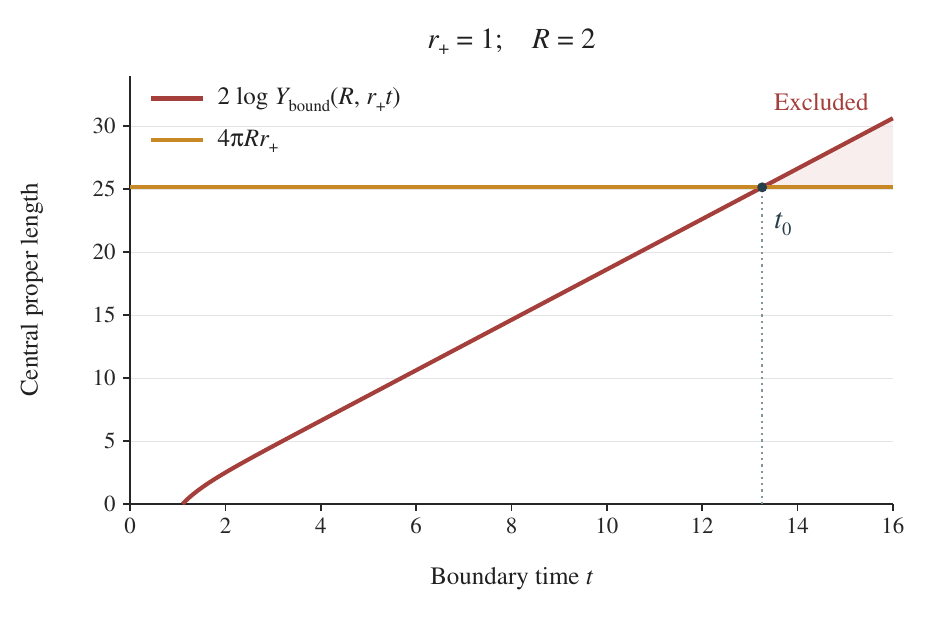}
\caption{Incompatible length bounds for a surviving two-sided
component, illustrated at $r_+=1$, $R=2$. Its required central
length grows with $t$, whereas recoloring costs at most $4\pi Rr_+$.
Their intersection is $t_0(R;r_+)$: beyond it, no such component can
satisfy both stationarity and slice minimality. This is an exclusion
bound, not the actual transition time of a selected saddle.}
\label{fig:length-bounds}
\end{figure}

\subsection{A finite upper bound on the probing time}
\label{sec:finite-bound}
For a network $N_*$ connecting the two boundaries to be selected by the maximin prescription, the lower bound on $\Lc(\Gamma)$ and the recoloring upper bound on $D_B$ must be compatible:
\begin{equation}
2\log Y_{\mathrm{bound}}(R,T)
\le \Lc(\Gamma)\le D_B\le4\pi R\rp.
\label{eq:master}
\end{equation}
Here the first inequality applies when $Y_{\mathrm{bound}}(R,T)>1$.
Thus, a surviving two-sided network must satisfy $Y_{\mathrm{bound}}(R,T)\le e^{2\pi R\rp}$.
Using the definition of $Y_{\mathrm{bound}}$
and the monotonicity of $\sinh$, this gives
$t\le t_0(R;\rp)$, where
\begin{equation}
t_0(R;\rp)=\frac1{\rp}\left[
\operatorname{arctanh}\frac1R+
\operatorname{arcsinh}
\left(\frac{e^{2\pi R\rp}}{\sqrt{R^2-1}}\right)
\right].
\label{eq:t0}
\end{equation}
For $t>t_0(R;\rp)$, we have
$Y_{\mathrm{bound}}(R,T)>e^{2\pi R\rp}>1$,
so the graph length bound applies and
excludes two-sided connections.
The argument also applies when cuts pass through
junctions or meet edges tangentially, by a small
perturbation of $R$ and continuity of $t_0$.
Since $R>1$ is arbitrary, we define
\begin{equation}
t_{\rm U}(\rp)=\inf_{R>1}t_0(R;\rp).
\label{eq:tU}
\end{equation}
For $t>t_{\rm U}(\rp)$, no network connecting the two boundaries can satisfy both stationarity and slice minimality.
Since neither length bound depends on the boundary partition or network topology, $t_{\rm U}(\rp)$ is a common upper bound for all allowed finite partitions.

\subsection{Optimizing the bound}
\label{sec:optimized-bound}
For fixed $\rp>0$, the function $t_0(R;\rp)$ diverges both as $R\to1^+$ and as $R\to\infty$.
It therefore attains a minimum at some finite $R_{\rm opt}>1$.
Furthermore, optimizing over $R$ yields a bound that remains finite uniformly over all horizon radii.
As shown in Appendix~\ref{app:optimization},
$t_{\rm U}(\rp)$ is strictly decreasing for
$\rp>0$ and satisfies
\begin{equation}
 \begin{gathered}
 2\pi<t_{\rm U}(\rp)<C_*,\\
 C_*=2\pi e^{k_*}=22.56368170\ldots.
 \end{gathered}
 \label{eq:temperature-ceiling}
\end{equation}
Here $k_*>1$ is the unique solution of
$k_*-1=e^{-k_*}$. The limits at
$\rp\to0$ and $\rp\to\infty$ are
$C_*$ and $2\pi$, respectively.
The optimized bound has the high-temperature expansion
\begin{equation}
\begin{aligned}
t_{\rm U}(\rp)
&=2\pi+\frac{1+\log(4\pi\rp)}{\rp}\\
&\quad+O\!\left(\rp^{-2}e^{-4\pi\rp}\right),\qquad \rp\to\infty.
\end{aligned}
\label{eq:bound-asymptotic}
\end{equation}
Appendix~\ref{app:optimization} gives an explicit remainder bound.
At fixed party number $\mathtt q$, Appendix~\ref{app:finite-q}
gives a sharper upper bound $t_{\rm U}^{(\mathtt q)}(\rp)$.
For fixed $\rp>0$, these bounds satisfy
\begin{equation}
 \begin{gathered}
 t_{\rm U}^{(\mathtt q)}(\rp)
 <t_{\rm U}^{(\mathtt q+1)}(\rp)
 <t_{\rm U}(\rp),\\
 \lim_{\mathtt q\to\infty}t_{\rm U}^{(\mathtt q)}(\rp)
 =t_{\rm U}(\rp).
 \end{gathered}
\end{equation}

\section{No late-time probing of the black-hole interior}
\label{sec:interior-exclusion}
For $t>t_{\rm U}(\rp)$, Sec.~\ref{sec:proof} shows that a stationary network selected by maximin cannot contain a connected component joining the two asymptotic boundaries.
Such a component would require a central interface length exceeding the recoloring bound.

The remaining components are those attached to only one boundary and those without boundary legs.
As shown in Appendix~\ref{app:localization}, the former cannot enter the black-hole interior $0<r<\rp$, while the latter lie on the bifurcation circle.
Consequently, no maximin network can probe the black-hole interior for $t>t_{\rm U}(\rp)$.

Defining $t_{\rm probe}(\mathcal P;\rp)$ as
the supremum of the times $t\ge0$ at which a maximin
network for a fixed boundary partition $\mathcal P$ can probe
this interior, with $t_{\rm probe}=0$ if no such time exists,
we obtain
\begin{equation}
 t_{\rm probe}(\mathcal P;\rp)
 \le t_{\rm U}(\rp)<C_*.
 \label{eq:uniform-probing-bound}
\end{equation}
Since this bound is independent of the number, sizes, and arrangement of the boundary parties, increasing the number of parties cannot extend the probing time indefinitely.
Equation~\eqref{eq:uniform-probing-bound} gives a common upper bound, not the exact time at which probing ceases for each partition; probing may cease earlier.

\section{Discussion}
\label{sec:discussion}
Previous work~\cite{Anegawa:2026yee} showed that genuine tripartite multi-entropy can probe Einstein--Rosen bridge growth for a finite period after bipartite entanglement entropies have saturated.
Our result establishes a universal limit on this probing time: 
For any fixed finite $\mathtt q$-partition of the two boundaries of nonrotating BTZ, the bulk network computing the leading classical multi-entropy $S^{(\mathtt q)}$ cannot have any interface that enters the black-hole interior once $t>t_{\rm U}(\rp)$.
Neither increasing the number of parties nor changing their sizes and arrangement can extend the probing time indefinitely. The same holds for genuine multi-entropy, a linear combination of multi-entropies.

Beyond time-dependent wormhole growth, multi-entropy can probe other features of bulk geometry, such as twist moduli~\cite{Anegawa:2026npk}.
It would be interesting to determine more generally which bulk geometric quantities multi-entropy can detect and how this sensitivity depends on the number and arrangement of the boundary parties.

Our conclusion relies on the multiway maximin prescription and the assumed spacetime stationarity of its selected networks.
Justifying the multiway maximin prescription and establishing the spacetime stationarity of the selected networks remain important open problems.
It would also be interesting to determine whether similar finite bounds hold in rotating BTZ and more general time-dependent geometries, and how quantum corrections modify the result.

\begin{center}
\textbf{Acknowledgments}
\end{center}
\begin{acknowledgments}
T.A.~is supported by JSPS KAKENHI Grant No.~26K17161. N.I.~was supported in part by  NSTC of Taiwan Grant Number 114-2112-M-007-025-MY3, and by MEXT KAKENHI Grant-in-Aid for Transformative Research Areas A “Extreme Universe” No.~21H05184. 
We used GPT-6 Astra (OpenAI) extensively to assist with analytical
calculations, the development and examination of mathematical
arguments, numerical computations, code development, and manuscript
preparation.
The authors directed the analysis, critically examined the mathematical
reasoning, and checked the results through analytical and numerical
consistency tests.
All scientific results and conclusions remain the responsibility
of the authors.
\end{acknowledgments}

\bibliographystyle{apsrev4-2}
\bibliography{reference.bib}

%\clearpage

%\vspace{1mm}

%\appendix
%\begin{center}
%\textbf{Appendix}
%\end{center}

%\vspace{-1mm}

\appendix

\vspace{-1mm}

\section{Embedding and length conventions}
\label{app:embedding}

Use $X\cdot X=-1$ in $\mathbb R^{2,2}$ with metric $\operatorname{diag}(-1,-1,1,1)$. In either exterior,
\begin{equation}
 \begin{gathered}
 X^0=\rho\cosh p,\qquad X^2=\rho\sinh p,\\
 X^1=\sqrt{\rho^2-1}\sinh(\rp t_\sigma),\\
 X^3=\sigma\sqrt{\rho^2-1}\cosh(\rp t_\sigma),\\
 \rho=r/\rp,\qquad p=\rp\phi\sim p+2\pi\rp.
 \end{gathered}
 \label{eq:embedding-chart}
\end{equation}
The quotient acts on $(X^0,X^2)$ only. Consequently $Y=X^1$ and $W=X^3$ are single valued even on winding edges. The same scalars continue through the interior with $Y^2-W^2=1-\rho^2$. At boundary time $T=\rp t$, define the null vector
\begin{equation}
 B_\sigma(p) =(\cosh p,\sinh T,\sinh p,\sigma\cosh T).
 \label{eq:boundary-vector}
\end{equation}
A point $X_b$ at the radial cutoff $r=r_b$, with the same time and angle, has embedding coordinates
\begin{equation}
 X_b \approx \frac{r_b}{\rp}B_\sigma(p) +O\!\left(\frac{\rp}{r_b}\right).
\end{equation}
Thus $B_\sigma(p)$ specifies the asymptotic direction
of the corresponding boundary point.

To describe a geodesic that winds around the angular direction, we replace the angular coordinate $\phi$ of one endpoint by $\phi+2\pi m$, or equivalently $p$ by $p+2\pi\rp m$, with $m\in\mathbb Z$, before evaluating the embedding inner products.

The length of an internal edge with endpoints $X,Z$ and
the subtracted length of a boundary leg ending at
$B=B_\sigma(p)$ are
\begin{equation}
 \begin{aligned}
 \ell(X,Z)&=\operatorname{arccosh}(-X\cdot Z),\\
 \widehat\ell(B,X)&=\log(-2B\cdot X).
 \end{aligned}
 \label{eq:embedding-lengths}
\end{equation}
These expressions require $-X\cdot Z>1$ and
$-2B\cdot X>0$, respectively.
Indeed, using the asymptotic form of $X_b$, the boundary-leg
length at the cutoff is
\begin{equation}
 \ell(X,X_b)
 =\log\frac{r_b}{\rp}+\widehat\ell(B,X)+o(1).
\end{equation}

Let $J_{\mathcal P}$ be the number of boundary points where the party label changes.
Each such point contributes one boundary leg. 
Subtracting the common divergent term for each leg, we define the renormalized network length by
\begin{equation}
 \Lhat(N)=\lim_{r_b\to\infty} \left[ \Lc(N)-J_{\mathcal P}\log\frac{r_b}{\rp} \right].
 \label{eq:subtraction}
\end{equation}
The subtraction is independent of time.

For a spacelike geodesic $X(s)$ parametrized by proper length $s$,
primes denote derivatives with respect to $s$.
To see why $X''$ must be proportional to $X$, consider a variation $\delta X(s)$ that keeps the curve on the AdS hyperboloid and fixes its endpoints. Preserving $X\cdot X=-1$ to first order requires $X\cdot\delta X=0$. Since the curve is a geodesic, its length is stationary under every such variation:
\begin{equation}
 \delta L
 =\int \frac{X'\cdot\delta X'}{\sqrt{X'\cdot X'}}\,ds
 =-\int X''\cdot\delta X\,ds=0.
\end{equation}
Here we used $X'\cdot X'=1$ on the original curve and integrated by parts; the endpoint term vanishes because the endpoints are fixed. Thus $X''$ has zero inner product with every allowed displacement $\delta X$. These displacements span the three-dimensional subspace defined by $X\cdot\delta X=0$, whose orthogonal complement is the one-dimensional space spanned by $X$. Hence $X''=\lambda X$.

\section{Length bound for branched networks}
\label{app:graph}
We prove the length bound~\eqref{eq:graph-lemma}
for general finite networks, including arbitrary
branches and cycles. We show the following.

{\it Claim}.\ Let $\Gamma$ be a finite connected graph with total length $D$. Each edge is parametrized by proper length $s$. 
Let $Y$ be continuous on $\Gamma$, satisfy $Y''=Y$ on every edge, and obey the Kirchhoff condition
\begin{equation}
\sum_{e\ni v}\partial_{n_e}Y=0
 \label{eq:graph-ode-kirchhoff}
\end{equation}
at every internal vertex $v$, where derivatives point into the incident edges. 
Loop incidences are counted separately. At every boundary vertex $b$, assume
\begin{equation}
 Y(b)\ge Y_{\rm bound}>1.
 \label{eq:graph-boundary-data}
\end{equation}
Suppose that a simple path $\gamma$ connecting
two distinct boundary vertices contains a point
$x_0$ with $Y(x_0)\le1$. Then
\begin{equation}
 D\ge2\log Y_{\rm bound}. 
 \label{eq:graph-bound-complete}
\end{equation}
In the BTZ application, the boundary vertices of $\Gamma$ are intersections with the cut circles $C_L$ and $C_R$, not the asymptotic boundaries.

{\it Proof}.\ We first establish $Y>0$ on $\Gamma$. 
Let $m$ be the minimum of $Y$ on $\Gamma$. If $m\le0$, the minimum cannot occur at a boundary vertex. 
At an edge-interior minimum, $Y'=0$ and $Y''=m\le0$, contradicting the minimum condition when $m<0$. If $m=0$,
uniqueness for $Y''=Y$ forces $Y=0$ along that edge.
At an internal vertex attaining $m$, all outward derivatives are nonnegative. 
Therefore, by the Kirchhoff condition~\eqref{eq:graph-ode-kirchhoff}, each derivative must vanish.
Along each incident edge, take $s=0$ at this vertex
and let $s$ increase into the edge. Then, the solution is $Y(s)=m\cosh s$. For $m<0$, this contradicts the definition of $m$ as the minimum. For $m=0$, the zero propagates through the connected graph to a boundary vertex, contradicting $Y \ge Y_{\rm bound}$. Hence $Y>0$.

Retain the part of $\Gamma$ where $Y \le Y_{\rm bound}$, subdividing edges at $Y=Y_{\rm bound}$, and denote the resulting graph by $K$.
Since $Y''=Y>0$, $Y$ is strictly convex on each edge and thus intersects the level $Y_{\rm bound}$ at most twice, so $K$ is finite, though not necessarily connected.
Starting from $x_0$ along $\gamma$ in both directions, let $p_-$ and $p_+$ be the first points where $Y =Y_{\rm bound}$.
They exist by the boundary conditions. The subpath $\gamma_s$ between them lies entirely in $K$, and therefore
\begin{equation}
 \Lc(\gamma_s)\le\Lc(K)\le D.
 \label{eq:graph-truncated-length}
\end{equation}
Since $Y > 0$, define $f=\log Y$. The edge
equation gives
\begin{equation}
 f''=1-(f')^2.
 \label{eq:graph-log-equation}
\end{equation}
At every internal vertex of $K$ with
$Y<Y_{\rm bound}$, the Kirchhoff condition remains
valid for $f$, since
\begin{equation}
 \sum_{e\ni v}\partial_{n_e}f(v)
 =\frac1{Y(v)}
   \sum_{e\ni v}\partial_{n_e}Y(v)=0.
\end{equation}
At an endpoint $v$ of an edge $e$, let
$\partial_{n_e}f(v)$ denote the derivative directed
from $v$ into the edge.
Thus, integrating along an edge gives
\[
 \int_e f''\,ds
 =-\sum_{v\in\partial e}\partial_{n_e}f(v).
\]
Summing over all edges of $K$, the contributions at internal
vertices cancel by the Kirchhoff conditions.
Using $f''=1-(f')^2$, we therefore obtain
\begin{equation}
 \int_K(f')^2\,ds
 =\Lc(K)+
 \sum_{v\in\partial K}
 \sum_{\substack{e\ni v\\e\subset K}}
 \partial_{n_e}f(v).
 \label{eq:graph-log-integral}
\end{equation}
The outer sum runs over all boundary vertices $v\in\partial K$. Since $v$ may coincide with an original junction, several edges may meet there.
The inner sum over $e\ni v$ includes all edges incident on $v$, but only those retained in $K$.
At each boundary vertex of $K$, we have $Y = Y_{\rm bound}$, while $Y \le Y_{\rm bound}$ along every incident edge retained in $K$. 
Hence $\partial_{n_e}f(v)\le0$. The internal vertex terms cancel, so
\begin{equation}
 \int_K(f')^2\,ds\le\Lc(K).
 \label{eq:graph-log-energy}
\end{equation}
At the two endpoints of $\gamma_s$,
$f(p_\pm)=\log Y_{\rm bound}$, whereas
$f(x_0)\le0$. Thus
\begin{equation}
 \int_{\gamma_s}|f'|\,ds
 \ge2[\log Y_{\rm bound}-f(x_0)]
 \ge2\log Y_{\rm bound}.
 \label{eq:graph-log-variation}
\end{equation}
On the other hand, Cauchy--Schwarz and
\eqref{eq:graph-log-energy} imply
\begin{equation}
 \begin{aligned}
 \left(\int_{\gamma_s}|f'|\ ds\right)^2
 &\le \left( \int_{\gamma_s}1^2\ ds \right)
\left( \int_{\gamma_s}(f')^2\ ds\right)\\
 &\le\Lc(\gamma_s)\Lc(K) \le\Lc(K)^2.
 \end{aligned}
 \label{eq:graph-log-CS}
\end{equation}
Combining these inequalities gives
\begin{equation}
 2\log Y_{\rm bound}\le\Lc(K)\le D,
 \label{eq:graph-limit-bound}
\end{equation}
which proves~\eqref{eq:graph-bound-complete}.
The path condition is essential: a low
point confined to a dangling branch does
not guarantee the two distinct transitions
from $Y\le1$ to $Y=Y_{\rm bound}$ used above.

\section{Localization of stationary components}
\label{app:localization}
We show that stationary network components attached to only one boundary, as well as components without boundary legs, cannot enter the open black-hole interior $0<r<\rp$. 
We use the single-valued embedding functions $Y=X^1$ and $W=X^3$, which satisfy the geodesic equations and junction balance conditions in Eq.~\eqref{eq:graph-equations}.

\subsection{Components attached to one boundary}
Consider a connected component $\Gamma$ attached only to boundary $\sigma=\pm1$, with all boundary legs anchored at $t_\sigma=t$. 
Set $T=\rp t$ and define a linear function of the embedding vector $X$ by
\begin{equation}
 f(X)=X^1\cosh T-\sigma X^3\sinh T.
 \label{eq:f-localization}
\end{equation}
Along each edge, $f(X(s))$ satisfies $f''=f$, where primes denote proper-length derivatives.
These derivatives also satisfy the Kirchhoff conditions at the junctions.

On each boundary leg, let the proper length $s$ increase toward infinity.
Since the geodesic equation is $X''=X$, its solution can be written as
\begin{equation}
 X(s)=Ae^s+Ce^{-s},
\end{equation}
where $A$ and $C$ are constant embedding vectors.
The growing term determines the asymptotic direction,
so $A\propto B_\sigma(p)$. By linearity,
\begin{equation}
 f(X(s))=f(A)e^s+f(C)e^{-s},
\end{equation}
where $f(A)=0$ follows from $A\propto B_\sigma(p)$
and $f(B_\sigma(p))=0$.
Thus, on every boundary leg of this one-sided component,
$f(X(s))=f(C)e^{-s}$ and $f'=-f$.
In particular, $f(X(s))\to0$ as $s\to\infty$.

If the component were instead attached to both boundaries,
then $f(B_{-\sigma}(p))=\sinh(2T)\ne0$ for $t>0$,
so the growing mode would remain on the boundary legs
attached to the opposite boundary.

For the one-sided component, using $f''=f$, we have
$(ff')'=(f')^2+f^2$.
Integrating edge by edge over $\Gamma$, the junction
terms cancel by the Kirchhoff conditions, leaving
\begin{equation}
 \int_\Gamma [(f')^2+f^2]\,ds =\sum_b f(b) f'(b) =-\sum_b f(b)^2.
 \label{eq:localization-energy}
\end{equation}
Here $b$ runs over the asymptotic endpoints of the boundary legs, and each boundary term is understood as a limit toward infinity.
Since $f$ tends to zero along every boundary leg of this one-sided component, the right-hand side vanishes.
The integrand is nonnegative, so $f=0$ throughout $\Gamma$.

Therefore,
$Y=\sigma W\tanh T$, and the embedding constraint
$Y^2-W^2=1-\rho^2$ gives
\begin{equation}
 \rho^2=1+\frac{W^2}{\cosh^2 T}\geq1.
 \label{eq:one-sided-localization}
\end{equation}
Thus no component attached to only one boundary
can enter the open horizon interior.

\subsection{Components without boundary legs}

For a connected component without boundary legs,
the same integration-by-parts argument applied to
$Y$ and $W$ has no boundary terms:
\begin{equation}
 \int_\Gamma \left[(Y')^2+Y^2+(W')^2+W^2\right]ds=0.
\end{equation}
Consequently, $Y=W=0$, and the embedding
constraint gives $\rho=1$. Such a component
lies on the bifurcation circle and does not
enter the open interior.

Therefore, once two-sided components are excluded
at $t>t_{\rm U}(\rp)$ by Sec.~\ref{sec:proof},
no maximin network can have an interface in
$0<r<\rp$.

\section{Optimization of the common bound}
\label{app:optimization}
We optimize the cut radius $R r_+$, determine the temperature dependence of the resulting bound, and sharpen the
recoloring estimate at fixed party number.

\subsection{A unique minimizing cut}
For convenience, set $b=2\pi\rp$, $k=bR>b$, and
$\Theta=\rp t_0(R;\rp)$.
Equation~\eqref{eq:t0} becomes
\begin{equation}
 \frac{k}{b}\sinh\Theta-\cosh\Theta=e^k.
 \label{eq:optimization-implicit}
\end{equation}
This equation implicitly determines $t_0$ as a function
of $b$ and $k$.
For fixed $b>0$, minimizing $t_0$ over $k>b$ gives
$t_{\rm U}(\rp)$.

Since $t_0$ diverges as $R\to1^+$ and $R\to\infty$,
its minimum is attained in the interior.
At a stationary cut, $d\Theta/dk=0$.
Differentiating Eq.~\eqref{eq:optimization-implicit}
with respect to $k$ and then substituting the result
back into the same equation gives
\begin{equation}
 \sinh\Theta=be^k,\qquad
 \cosh\Theta=(k-1)e^k.
 \label{eq:optimal-stationarity}
\end{equation}
Eliminating $\Theta$, we find that the minimizing value
$k(b)$ must satisfy
\begin{equation}
 b^2=(k-1)^2-e^{-2k}.
 \label{eq:optimal-root}
\end{equation}
Since $\cosh\Theta>0$, we require $k>1$.

The right-hand side of Eq.~\eqref{eq:optimal-root}
is strictly increasing on $k>1$, from a negative value
at $k=1$ to infinity.
Thus, for each $b>0$, there is a unique root $k(b)>1$.
This root satisfies $k>b+1$ and therefore lies in
the allowed domain $k>b$.
Since a minimum exists and every stationary point must
satisfy Eq.~\eqref{eq:optimal-root}, this unique root gives
the global minimum:
\begin{equation}
 R_{\rm opt}=\frac{k(b)}{b},\quad  t_{\rm U}(\rp) =\frac{2\pi}{b} \operatorname{arcsinh}\!\left(be^{k(b)}\right).
 \label{eq:optimized-bound-exact}
\end{equation}

At the minimizing cut, we write
\begin{equation}
 \Theta(b)=\rp t_{\rm U}(\rp)
 =\operatorname{arcsinh}\!\left(be^{k(b)}\right).
\end{equation}
As $b\to0$, Eq.~\eqref{eq:optimal-root} implies that $k(b)$ approaches the unique finite root $k_*>1$ satisfying
\begin{equation}
 k_*-1=e^{-k_*}.
\end{equation}
Consequently, $\Theta(b)\to0$. We therefore extend $\Theta$ continuously to $b=0$ by setting $\Theta(0)=0$.

\subsection{Monotonicity and the low-temperature limit}
We now examine how the optimized bound $t_{\rm U}(\rp)$ varies with $\rp$, or equivalently with $b$.
Differentiating Eq.~\eqref{eq:optimal-root} and the optimized expression~\eqref{eq:optimized-bound-exact} with respect to $b$ gives
\begin{equation}
 \begin{aligned}
 k'(b)&=\frac{b}{k-1+e^{-2k}}>0,\\
 \Theta'(b)&=\frac{k}{k-1+e^{-2k}}.
 \end{aligned}
 \label{eq:optimal-derivatives}
\end{equation}
The second expression decreases with $k$, since
\begin{equation}
 \frac{d}{dk}\frac{k}{k-1+e^{-2k}}
 =\frac{(1+2k)e^{-2k}-1}
 {(k-1+e^{-2k})^2}<0.
\end{equation}
Together with $k'(b)>0$, this implies $\Theta''(b)<0$.
Thus $\Theta'$ is strictly decreasing on $(0,\infty)$.
Therefore, for any fixed $b>0$, we have $\Theta'(s)>\Theta'(b)$ whenever $0<s<b$.
Using $\Theta(0)=0$, we therefore obtain
\begin{equation}
\begin{aligned}
 \frac{d}{db}\left(\frac{\Theta(b)}{b}\right)
 &=\frac{b\Theta'(b)-\Theta(b)}{b^2}\\
 &=\frac{1}{b^2}\int_0^b
 [\Theta'(b)-\Theta'(s)]\,ds<0.
\end{aligned}
\end{equation}
Thus $t_{\rm U}(\rp)=2\pi\Theta(b)/b$ is strictly decreasing with $\rp$.

Its supremum over $\rp>0$ is therefore given by
the limit $b\to0$.
Using $k(b)\to k_*$ and $\operatorname{arcsinh}x\sim x$
as $x\to0$, we find
\begin{equation}
 \begin{aligned}
 k_*&=1.278464542761\ldots,\\
 \lim_{\rp \to 0}t_{\rm U}(\rp) &=2\pi e^{k_*} =22.563681698501\ldots.
 \end{aligned}
\end{equation}
Although $R_{\rm opt}$ diverges in this limit, the physical cut radius $R_{\rm opt}\rp=k(b)/(2\pi)$ remains finite.

\subsection{High-temperature expansion with a remainder bound}
Equation~\eqref{eq:t0} can be written exactly as
\begin{equation}
\begin{aligned}
\rp t_0(R;\rp)
={}&bR-\log(R-1)\\
&+\log\left[
1+\sqrt{1+(R^2-1)e^{-2bR}}
\right].
\end{aligned}
\end{equation}
The last logarithm is at least $\log2$.
Minimizing the remaining expression over
$R>1$ gives
\begin{equation}
 t_{\rm U}(\rp)\ge B(\rp),\qquad
 B(\rp)=2\pi+\frac{1+\log(4\pi\rp)}{\rp},
\end{equation}
with the minimum attained at $R_0=1+1/b$.

Evaluating $t_0$ at this same $R_0$,
and using $\sqrt{1+x}\le1+x/2$
and $\log(1+x)\le x$, yields
\begin{equation}
 0\le t_{\rm U}(\rp)-B(\rp)
 \le\frac{2/b+1/b^2}{4\rp}e^{-2b-2}.
 \label{eq:explicit-asymptotic-error}
\end{equation}
This proves Eq.~\eqref{eq:bound-asymptotic} and $\lim_{\rp\to\infty}t_{\rm U}(\rp)=2\pi$.
Together with strict monotonicity and the low-temperature limit, this establishes Eq.~\eqref{eq:temperature-ceiling}.

\subsection{A sharper bound at fixed party number}
\label{app:finite-q}

For a partition with $\mathtt q$ parties,
the recoloring cost can be reduced by choosing
the label appropriately. Let
$L_C=\Lc(C_L)+\Lc(C_R)$, and let $\ell_i$
be the total length of the portions of these
cuts whose label immediately outside $B$
is $i$, summed over both cuts.
Then $\sum_{i=1}^{\mathtt q}\ell_i=L_C$,
so some label occupies length at least
$L_C/\mathtt q$. Recoloring $B$ with the
label of largest $\ell_i$ gives
\begin{equation}
 \begin{aligned}
 D_B
 &\le L_C-\max_{1\le i\le\mathtt q}\ell_i\le\left(1-\frac1{\mathtt q}\right)L_C\\
 &\le4\pi R\rp \left(1-\frac1{\mathtt q}\right).
 \end{aligned}
 \label{eq:central-upper-q}
\end{equation}
This argument allows each label to occupy
multiple disconnected portions of either cut.

Set $\alpha_{\mathtt q}=1-1/\mathtt q$.
Combining~\eqref{eq:central-upper-q}
with~\eqref{eq:graph-lemma}, a surviving
two-sided component must satisfy
\begin{equation}
 Y_{\mathrm{bound}}(R,\rp t)
 \le e^{2\pi\alpha_{\mathtt q}R\rp}.
 \label{eq:central-condition-q}
\end{equation}
Solving the equality in~\eqref{eq:central-condition-q}
for $t$ gives
\begin{equation}
 t_0^{(\mathtt q)}(R;\rp)
 =\frac{1}{\rp}\left[
 \operatorname{arctanh}\frac{1}{R}
 +\operatorname{arcsinh}
 \left(\frac{e^{2\pi\alpha_{\mathtt q}R\rp}}{\sqrt{R^2-1}}\right)
 \right].
\end{equation}
Comparing this expression with Eq.~\eqref{eq:t0},
we find
$t_0^{(\mathtt q)}(R;\rp)
=\alpha_{\mathtt q}t_0(R;\alpha_{\mathtt q}\rp)$:
replacing $\rp$ by $\alpha_{\mathtt q}\rp$ in
$t_0$ gives the same expression inside the brackets,
while the prefactor is restored by multiplying by
$\alpha_{\mathtt q}$.
Since $\alpha_{\mathtt q}>0$ is independent of $R$,
minimizing over $R>1$ therefore yields
\begin{equation}
 \begin{aligned}
 t_{\rm U}^{(\mathtt q)}(\rp) \equiv\inf_{R>1}t_0^{(\mathtt q)}(R;\rp) =\alpha_{\mathtt q} t_{\rm U}(\alpha_{\mathtt q}\rp).
 \end{aligned}
 \label{eq:finite-q-time}
\end{equation}
At $t=t_0^{(\mathtt q)}(R;\rp)$, we have $Y_{\mathrm{bound}}=e^{2\pi\alpha_{\mathtt q}R\rp}>1$.
Since $Y_{\mathrm{bound}}$ increases with $t$, the condition $Y_{\mathrm{bound}}>1$ also holds for all later times.

Together with the localization result in
Appendix~\ref{app:localization}, this improved
bound gives
$t_{\rm probe}(\mathcal P;\rp)
\le t_{\rm U}^{(\mathtt q)}(\rp)$
for any fixed partition with $\mathtt q$ parties.
Equation~\eqref{eq:temperature-ceiling}
further gives
\begin{equation}
 2\pi\alpha_{\mathtt q}
 <t_{\rm U}^{(\mathtt q)}(\rp)
 <\alpha_{\mathtt q}C_*.
 \label{eq:finite-q-ceiling}
\end{equation}
Thus finite party number sharpens the bound.
Writing $b=2\pi\rp$, Eq.~\eqref{eq:finite-q-time} gives
\begin{equation}
 t_{\rm U}^{(\mathtt q)}(\rp)
 =\frac{2\pi}{b}\Theta(\alpha_{\mathtt q}b).
\end{equation}
Since $\Theta'>0$ by Eq.~\eqref{eq:optimal-derivatives}
and $\alpha_{\mathtt q}<\alpha_{\mathtt q+1}<1$,
this bound is strictly increasing in $\mathtt q$
and remains below $t_{\rm U}(\rp)$.
By continuity of $\Theta$, it converges to
$t_{\rm U}(\rp)$ as $\mathtt q\to\infty$.

\end{document}